# Feasibility Study of Gas Electron Multiplier Detector as an X-Ray Image Sensor

Sukyoung Shin, Jaehoon Jung, and Soonhyouk Lee*

*Ewha Medical Research Institute, School of Medicine, Ewha Womans University, Seoul 158-710, Korea*

For its ease manufacturing, flexible geometry, and cheap manufacturing cost, the gas electron multiplier (GEM) detector can be used as an x-ray image sensor. For this purpose, we acquired relative detection efficiencies and suggested a method to increase the detection efficiency in order to study the possibility of GEM detector as an x-ray image sensor.

The GEM detector system is composed of GEM foils, the instrument system, the gas system, and the negative power supply. The instrument system consists of the A225 charge sensitive preamp, A206 discriminator, and MCA8000D multichannel analyzer. For the gas system, Argon gas was mixed with $CO_2$ to the ratio of 8:2, and for the negative 2,000 volts, the 3106D power supply was used. The CsI-coated GEM foil was used to increase the detection efficiency.

Fe-55 was used as an x-ray source and the relative efficiency was acquired by using the ratio of GEM detector to the CdTe detector. The total count method and the energy spectrum method were used to calculate the relative efficiency. The relative detection efficiency of GEM detector for Fe-55 by using total count method was 32 % and by using energy spectrum method, the relative efficiencies were 5, 43, 33, 37, 35, and 36 % respectively according to the energy spectrum of 2, 3, 4, 5, 6, and 7 KeV.

In conclusion, we found that the detection efficiency of the two layered GEM detector is insufficient for the x-ray image sensor, so we suggested a CsI coated GEM foil to increase the efficiency rate and the result value was increased to 41 %.

Email: king6d@ewha.ac.kr

Fax: +82-2-2654-0363

# I. INTRODUCTION

Gas Electron Multiplier (GEM) detector is one of the gaseous ionization detector which is primordially suggested by Fabio Sauli of CERN. GEM foil consists of many small holes which can act as an individual multiplier. When high voltage is applied to GEM foil, inhomogeneous electric field can be generated inside holes and drift electron transferred to the hole is multiplied by Avalanche effect according to the intensified electric field [1-2].

The Avalanche can be generated on the GEM foil as thin as 70 µm. The flexibility of the size of GEM foil and the cheap manufacturing cost make GEM detector useful to be applied to various applications such as radiation detection and medical radiation imaging system [3-6].

In order to use the GEM detector as an imaging system, the quality of the GEM detector should be investigated. One of the most important factors in determination of quality of GEM detector is effective gain acquired by electron multiplication. Many investigations were implemented to analyze the geometrical and physical factors to increase the effective gain of GEM detector [1-3,7]. In addition to the effective gain, efficiency is also an important factor to determine the validation of GEM detector in order to be used in various applications [8-10].

The efficiency of GEM detector can be influenced by the voltage between GEM foils, collection electric field, drift electric field, combined gas ratio, number of GEM foils, and the geometry of GEM detector. In order to minimize these effects, it is useful to use relative efficiency to the well-known radiation detector [10]. In the previous study [10], we acquired the relative efficiency with the total count method, but it might lose information of energy spectrum. Therefore, both the total count method and the energy spectrum method for the relative efficiency are required.

In this study, we acquire the relative detection efficiency by the total count method and the energy spectrum method in order to investigate the feasibility of GEM detector as an x-ray imaging sensor. Also, we suggest a method to increase the detection efficiency by using CsI-coated GEM detector.

## II. METHOD AND MATERIAL

### A. Development of GEM detector system

The GEM detector system was composed as in the previous experiment [10]. It consists of GEM detector, instrumentation system, negative high voltage supply, and gas circulation system. For the instrumentation system, A225 (Amptek, USA) charge sensitive preamp and A206 (Amptek, USA) discriminator were used to make pulse signal. The pulse signal is, then, AD converted and counted by MCA8000D (Amptek, USA) multichannel analyzer and the spectrum of the signal is extracted by DppMCA software (ver.4.3.1, Amptek, USA). The 3106D (Canberra, USA) high voltage supply was used to apply negative 2,000 V to the GEM detector. Mixed gas consisted of Argon and $CO_2$ with the ratio of 80 to 20 was supplied to the GEM detector and exhausted through bubbler.

Fig. 1 shows the GEM detector and Fig.2 shows the structure of double GEM layers. Two GEM foils which were purchased from the CERN were employed to compose the double GEM detector. The distance of drift, transfer, and collection region are 4, 2, and 1 mm respectively and negative 400 V was applied to both GEM foils. The effective area for the radioactive irradiation is 2.5 cm × 2.5 and the detection window consists of 70 μm thick copper cathode surrounded by kapton as thick as 65 μm.

### B. The relative detection efficiency

The relative detection efficiency was acquired by using the X-100R CdTe detector (Amptek, USA) and PX4 (Amptek, USA) data acquisition module. The the number of pulses from the radioactive source were counted. The efficiency of CdTe detector was provided by the manufacturer.

The detection efficiency of the GEM detector using total count method was acquired by Eq. 1 [10,11] The total efficiency was acquired by the ratio of the number of photon counts $(S(E))$ from the radiation source to the number of spectrum $(N_{Total}(E))$ measured [10,11].

$$\epsilon_{Total}(E) = \frac{N_{Total}(E)}{S(E)} \qquad (1)$$

In order to diminish the structural effect on the number of photon, Eq. 1 was modified to Eq.2 for the relative efficiency calculation.

$$\epsilon_{Total}(E_s) = \left( \frac{\frac{N_{Total}(E)_{gem}}{S(E)_{gem}}}{\frac{N_{Total}(E)_{CdTe}}{S(E)_{CdTe}}} \right) \times \epsilon_{CdTe}(E_s) = \frac{N_{Total}(E)_{gem}}{N_{Total}(E)_{CdTe}} \times \epsilon_{CdTe}(E_s) \quad (2)$$

where $\epsilon_{Total}(E_s)$ is a relative total efficiency for the measured energy spectrum, $N_{Total}(E)_{gem}$ is the total number of spectrum measured in the GEM, $N_{Total}(E)_{CdTe}$ is the total number of spectrum measured in the CdTe detector, and $\epsilon_{CdTe}(E_s)$ is efficiency of CdTe detector for the specific energy to be acquired.

From this total count method, relative efficiency was acquired 10 times in every 1,200 sec and averaged.

The detection efficiency of the GEM detector for the energy spectrum method was acquired by Eq. 3, which was used in the previous research [6-7], except that the range of total count is divided into small region, *m*, the resolution of energy spectrum.

$$\left( N_m(Gem) \big/ N_m(CdTe) \right) \times E_m(CdTe), \quad (3)$$

where $N_m(Gem)$ and $N_m(CdTe)$ are the number of pulse counts of GEM detector and CdTd detector respectively and $E_m$ *(CdTe)* is the efficiency of CdTe detector. The resolution of energy spectrum *m* was 1 Kev.

### C. Radiation source

Disc-shaped Fe-55 (Spectrum Techniques, USA) was used for the experiment. Fe-55 has 5.9 keV peak, 25 μCi of activity, and 2.73 year of half-life. Radioactive source is located 0.5 cm above the detection windows of both detectors. Data is averaged after 10 measurements and the 1 measurement lasts for 10 seconds. Energy below 2 KeV was considered noise and the energy spectrum range of the experiment was set from 2 KeV to 7 KeV.

### D. CsI-coating

Fig. 3 shows the CsI-coated GEM foil which was purchased from CERN. The Csi coating was done under vacuum by PVD deposition. The thickness on the GEM foil is 300 nm.

### III. RESULTS AND DISCUSSION

Fig. 3 shows the measured number of counts from the Fe-55 by using GEM detector (Fig. 3(a)) and CdTe detector (Fi. 3(b)) according to the energy spectrum and the energy peak around 5.9 KeV and the argon escape peak were shown.

Table 1 shows the relative detection efficiency of Fe-55 acquired by the total count method from the data acquired in the Fig. 3. The measured net total count with the CdTe detector was 109,794 $\pm$ 3,385 and the measured net total count acquired with GEM detector was 39,027 $\pm$ 352. Therefore, the relative detection efficiency of the GEM detector to the CdTe detector by the total count method was around 32%.

Fig. 4 shows the relative efficiency of the GEM detector to the CdTe detector according to the Fe-55 energy spectrum. For the 6 ranges of the energy spectrum, that is, 2, 3, 4, 5, 6, and 7 KeV, the relative detection efficiency were 5, 43, 33, 37, 35, and 36 % respectively. The average relative efficiency acquired by the energy spectrum is 32 %. This average value is the same as that of the total count method. The difference between the average efficiency value and the individual efficiency value of each energy spectrum is due to the number detection count for the energy spectrum of 2 KeV and 3 KeV, as shown in the Fig. 4. They were ignored and averaged out in the total count method.

This relative detection efficiency of the double GEM detector for the Fe-55 shows that the detection efficiency is somewhat low to be used as an x-ray sensor [12]. In order to increase the detection efficiency, a variety of methods can be considered. For example, increase the amplification rate of the instrument, increase the number of the GEM layer [13], or increase the voltage across the GEM foils [1,14]. However, increasing the amplification rate can cause the decrease of the signal-to-noise ratio, and as a result, can cause decrease the detection rate. Increasing the number of GEM layer can cause saturation at the result and increasing the voltage across the GEM foil is subject to damage GEM foil [1].

Therefore, in this study, we suggested CsI-coated GEM foil in order to increase the detection efficiency. CsI is usually used as a scintillator, the photon converter. However, the light produced from the scintillator by radiation is usually detected by imaging sensor such as CCD or CMOS [15,16] and increase the cost of manufacturing. In order to solve this problem, we coated GEM foils with CsI as shown in the Fig. 3. The CsI-coated GEM foil produced photons and electrons by radiation, and the electrons were amplified in the next GEM layer. This method noes not need any light-sensitive sensor such as CCD or CMOS and can be used in the normal GEM detector structure, just only to replace a normal GEM foil with CsI-coated Gem foil. As a result acquired increased relative detection efficiency of 41 %.

## IV. CONCLUSION

In this study, we investigated the feasibility of the GEM detector as an x-ray imaging sensor by acquiring the relative detection efficiency. The detection efficiency of the GEM detector was lower than that of the CdTe detector, and the result showed that some methods to increase the detection efficiency might be required for the GEM detector to be used as an imaging sensor. The CsI-coated GEM foil is one of the methods and it showed that it increased the detection efficiency.

ACKNOWLEDGEMENT

This paper was supported by RP-Grant 2013 of Ewha Womans University and was supported by the Nuclear Power Core Technology Development Program of the Korea Institute of Energy Technology Evaluation and Planning (KETEP), granted financial resource from the Ministry of Trade, Industry & Energy, Republic of Korea. (No. 20131510400050)

REREFENCES

[1] F. Sauli, Nucl. Inst. Meth. A **386**, 531 (1997).

[2] F. Sauli, Nucl. Inst. Meth. **513,** 115 (2001).

[3] A. Buzulutskov, Inst. Exp. Tech. **50**, 287 (2007).

[4] T. Uchida et al, IEEE NSS/MIC, 1450 (2010).

[5] T. Koike et al, 2nd INTERNATIONAL CONFERENCE ON MICRO PATTERN GASEOUS DETECTORS, 1 (2012).

[6] S. Kang et al, J. Nucl. Sci. Tech. **4**, 330 (2004).

[7] G. Bencivenni et al, Nuc. Inst. Meth. In Phys. Res. A **494**, 156 (2002).

[8] V. Peskov *et al.*, IEEE Trans. Nucl. Sci. **48**, 1070 (2001).

[9] S.Y. Ha *et al.*, J. Korean Phys. **55**, 2366 (2009).

[10] S.Lee, J. Jung, and R.Lee, Korean J. Med. Phys. 25, 95 (2014).

[11] L.M.Christine, *Detection efficiency* (IAEA-ALMERA Technical Visit, 2010).

[12] M. W. Bautza et al, High-Energy Detectors in Astronomy, **5501**, 111 (2004).

[13] N. Abgrall, Master Thesis, Departement de Physique Nucleaire et Corpusculaire, Universite de Geneve (2006).

[14] J. Rzadkiewicz et al, Proceedings of the 2nd International Conference Frontiers in Diagnostic Technologies (2011).

[15] F. A. F. Fraga et al., Nucl. Inst. Meth. A525, 57(2004).

[16] T. Meinschad, L. Ropelewski, and F. Sauli, Nucl. Inst. Meth. A547, 342 (2005).

**Figure Captions.**

Fig. 1. GEM detector

Fig. 2. Double GEM detector structure.

Fig. 3. Csi-coated GEM foil.

Fig. 4. The number counts of Fe-55 energy spectrum:  (a) GEM Detector and  (b) CdTe Detector.

Fig. 5. Detection efficiency of GEM detector and CdTe detector.

Table 1. The relative efficiency of Fe-55

| CdTe* | GEM | | CsI-coated GEM | |
|---|---|---|---|---|
| Count | Count | Relative Efficiency (%) | Count | Relative Efficiency (%) |
| 109,794 ± 3,385 | 39,027 ± 352 | 32±0.01 | 50,120±475 | 41±0.02 |

Average: Mean ±S.D.

* : Detection efficiency of CdTe at 5.9 KeV: 90%

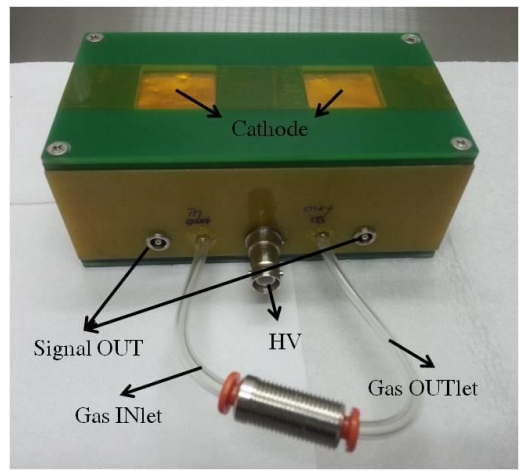

Fig. 1.

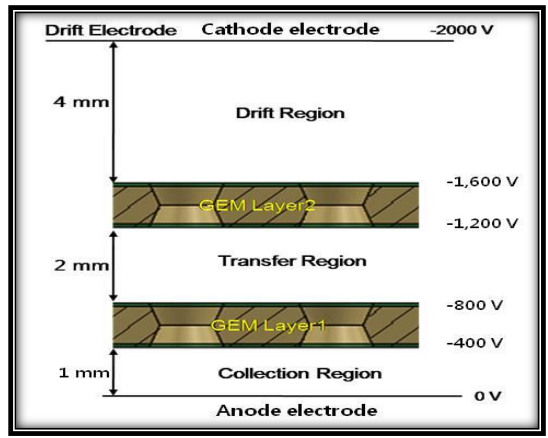

Fig. 2.

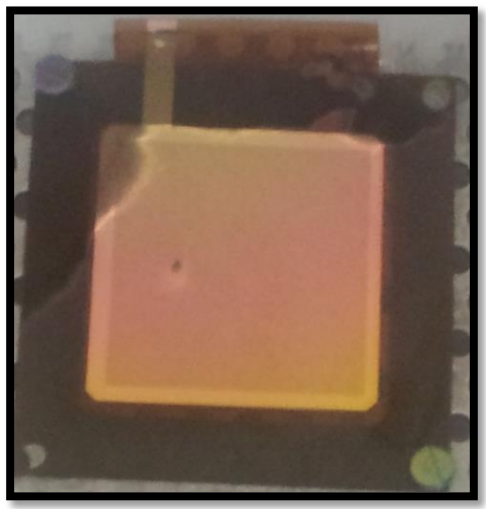

Fig. 3.

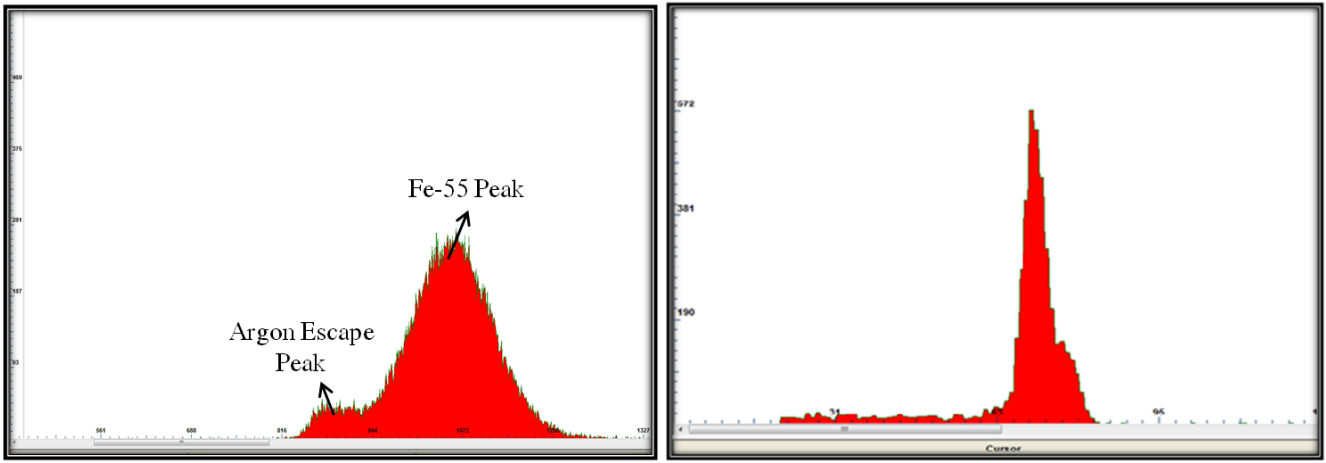

(a) (b)

Fig. 4.

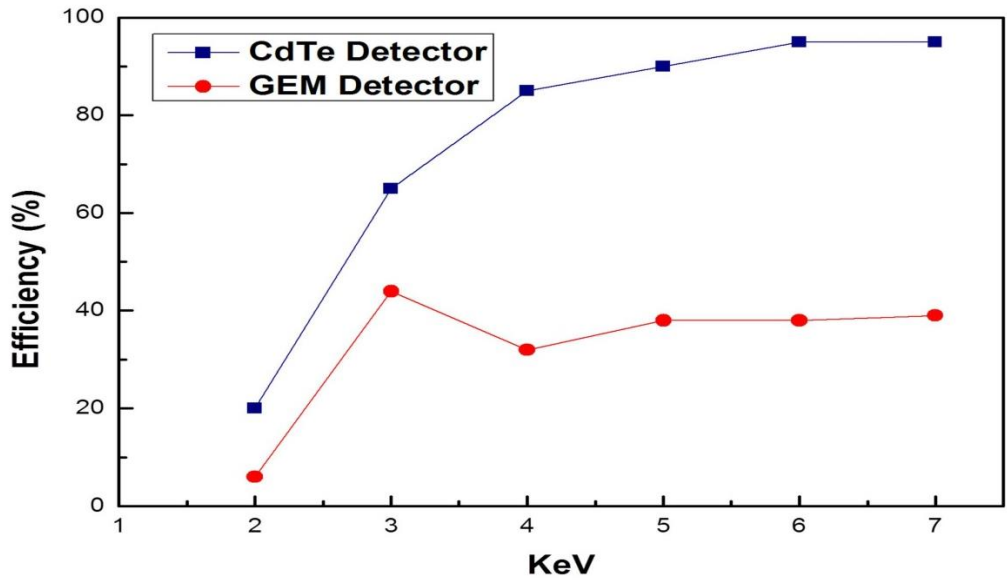

Fig. 5.